\documentclass[pra,aps,amsmath,twocolumn,showpacs,amsfonts,amssymb,floatfix,superscriptaddress]{revtex4-1}
\usepackage{graphicx}
\usepackage{grffile}
\usepackage{multirow}

\usepackage{times}
\usepackage[utf8]{inputenc}
\usepackage[english]{babel}
\usepackage{fancyhdr}

\usepackage{float}

\newcommand{\abs}[1]{\vert #1\vert}

\begin{document}

\title{Casimir interaction between inclined metallic cylinders}

\author{Pablo Rodriguez-Lopez}
\affiliation{Departamento de F\'\i sica Aplicada I and GISC,
Facultad de Ciencias F\'\i sicas, Universidad Complutense, 28040 Madrid, Spain}

\affiliation{Departamento de Matem\'aticas and GISC, Universidad Carlos III de Madrid,
Avenida de la Universidad 30, 28911 Legan\'es, Spain}

\author{Thorsten Emig}
\affiliation{Laboratoire de Physique Th\'eorique et Mod\`eles
  Statistiques, CNRS UMR 8626, B\^at.~100, Universit\'e Paris-Sud, 91405
  Orsay cedex, France}

\begin{abstract} 
The Casimir interaction between one-dimensional metallic objects (cylinders, wires) displays unconventional features. Here we study the orientation dependence of this interaction by computing the Casimir energy between two inclined cylinders over a wide range of separations. We consider Dirichlet, Neumann and perfect metal boundary conditions, both at zero temperature and in the classical high temperature limit. For all types of boundary conditions, we find that at large distances the interaction decays slowly with distance, similarly to the case of parallel cylinders, and at small distances scales as the interaction of two spheres (but with different numerical coefficients). 
Our numerical results at intermediate distances agree with our analytic predictions at small and large separations. Experimental implications are discussed.
\end{abstract}


\maketitle

\section{Introduction}

Collective phenomena in charge density and current fluctuations are known to generate Casimir-Lifshitz interactions \cite{Casimir_pp,Lifshitz} that are unconventionally long-ranged for one-dimensional metallic objects as cylinders or wires \cite{Emig+2006,Rahi-Cylinders}. These interactions attract currently substantial interest since Casimir forces could play an important role for micro and nano devices \cite{Ardito}.  In general, the building elements of such devices cannot be modeled as planar surfaces (as in Ref.~\onlinecite{Casimir_pp,Lifshitz}); more common shapes are one-dimensional structures as wires or beams. Interactions of Casimir or (non-retarded) van der Waals type between quasi one-dimensional shapes are also important in biological systems where rodlike particles auch as DNA, viruses, or microtuboles interact due to correlated charge fluctuations \cite{Parsegian-book,PSA-non-parallel-cylinders-salt-suspenssions}. Due to correlation effects that are particularly strong for conducting objects, the interaction is not properly 
described by pairwise summation approaches. For non-conducting objects, these summation schemes usually
yield the correct scaling of the interaction energy with distance and dimension of the objects. However, for metals the scaling is different and can be rather sensitive to the description of the material properties \cite{Barash-and-Kyasov,Noruzifar+2011}.
Proximity force approximations (PFA) are restricted to short separations  \cite{Review_Casimir} but for cylindrical shapes it is important to study interactions also at larger separations due to their slow decay. 

A recently developed scattering approach can be used to compute the interaction between various shapes, including cylinders, over a large range of separations \cite{EGJK,Review-Jamal-Emig}. In the following we shall employ the latter technique to study the orientation dependence of the Casimir interaction between metallic cylinders, both for a scalar field obeying Dirichlet or Neumann boundary conditions and for the electromagnetic field with perfect metal  boundary conditions. The temperature $T$ is assumed be either zero or in the classical limit  $k_B T \gg \hbar c$.  An understanding of the sensitivity to orientation is important since the measurement of cylinder interactions can be hampered by deviations from parallelism due to external perturbations. Orientation dependence plays also a role in mixtures of cylindrical shapes as carbon nanotubes and of the biological objects mentioned before where entropic effects compete with the direct interaction.  In the limit where the length of the objects is much smaller than their distance, the orientation dependence has been studied for metallic spheroids \cite{Spheroids_Emig}. Here we are interested in the opposite limit where the decay of the interaction is expected to be much slower \cite{Rahi-Cylinders,Rahi-Cylinders2}.  The case of non-parallel, infinitely long cylinders has been studied by making additivity assumptions \cite{PSA-non-parallel-cylinders-salt-suspenssions}. Recently, the van der Waals interaction between crossed cylinders, ignoring retardation effects, has been computed in the asymptotic large distance limit \cite{Dobson+2009}.  The force between inclined cylinders has been derived from a dilution process for anisotropic dielectric media \cite{Rajter+2007} and the results resemble pair-wise summation results. Advanced numerical techniques have been used to study the case of two perpendicular cylinders of finite length $L$ (capsules) \cite{Numerical-Casimir}. When $L$ is increased, the energy is found to approach an $L$-independent value. Our numerical results for sufficiently small distances (where a finite $L$ becomes less important) are consistent with the reported values.

The Casimir force between crossed metal cylinders has been measured \cite{Ederth2000}. More recently, Decca {\it et al.} discussed the possibility to measure the thermal Casimir force and its gradient between a plate and a microfabricated cylinder attached to a micromachined oscillator \cite{Decca+2010}. 

Before we consider the interaction at arbitrary separations, it is instructive to review the prediction of the PFA.
This approximation is often applied to describe the interaction at 
surface-to-surface distances $l=d-2R$ much shorter than the radii of curvature. For parallel cylinders of length $L\to \infty$, this approximation
yields at zero temperature  \cite{Rahi-Cylinders2}
\begin{equation}\label{PFA_Energy_T_0_parallel}
E_{0,\parallel}^{PFA} = - \frac{\pi^{3}}{1920}\hbar c L\sqrt{\frac{R}{l^{5}}} \, ,
\end{equation}
and in the high temperature limit
\begin{equation}\label{PFA_Energy_T_cl_parallel}
E_{cl,\parallel}^{PFA} = - \frac{\zeta(3)}{16}k_{B}T L\sqrt{\frac{R}{l^{3}}} 
\end{equation}
so that one has a finite energy per length in the limit of infinitely long cylinders.
For inclined cylinders with inclination angle $\theta\in [0,\pi/2]$, cf.~Fig.~\ref{fig:geometry}, we obtain (see App.~\ref{App. B}) at zero temperature
\begin{equation}\label{PFA_Energy_T_0_non_parallel}
E_{0,\theta}^{PFA} = - \frac{\pi^{3}}{720}\frac{\hbar c}{\sin(\theta)}\frac{R}{l^{2}} \, ,
\end{equation}
and in the high temperature limit
\begin{equation}\label{PFA_Energy_T_cl_non_parallel}
E_{cl,\theta}^{PFA} = - \frac{\zeta(3)}{4}\frac{k_{B}T}{\sin(\theta)}\frac{R}{l} \, .
\end{equation}
The energy is no longer extensive in $L$ but has a simple orientation dependence with a divergence $\sim 1/\theta$ for the approach of the parallel configuration. In the limit of close approach, $l\to 0$, the energy of inclined cylinders is less divergent in $l$ than in the parallel setup. For inclined cylinders at short distance one would expect a scaling of the PFA energy that resembles the one for two compact objects as, e.g., two spheres. Indeed, the scaling of the PFA energies of two spheres with $l$ and $R$ is the same as in Eqs.~\eqref{PFA_Energy_T_0_non_parallel}, \eqref{PFA_Energy_T_cl_non_parallel}. In the following, we shall describe how the interaction at larger separations deviates from this simple PFA estimates. We find that the interactions are no longer simple power-laws in $l$ but acquire logarithmic factors and that for inclined cylinders the decay at very large $l \gg R$ is even {\it slower} than for parallel cylinders.


The rest of this work is organized as follows.  In Sec.~\ref{sec: 2} we describe the system and review briefly the scattering approach that we use in the following sections and derive the so-called translation matrices that couple multipole moments of inclined cylinders.  In the following Sec.~\ref{sec: 4} we obtain analytic results at asymptotically large separations while in Sec.~\ref{sec: 6} we compute numerically the interaction at intermediate distances and compare it to PFA predictions. Finally, we present a conclusion and discussion of experimental implications in Sec.~\ref{sec: 7}. Mathematical details of the derivation of translation matrices and the PFA formulae are provided in App.~\ref{App. A} and App.~\ref{App. B}, respectively.

\section{Scattering approach}
\label{sec: 2}

\subsection{Geometry and Casimir energy}

Recently, a scattering approach for Casimir interactions between non-planar objects has been developed \cite{EGJK,Review-Jamal-Emig}. It relates the Casimir energy to the electromagnetic scattering properties of each object. The Casimir free energy at temperature $T$ can be written as 
\begin{equation}\label{Energy_T_finite}
E = k_{B}T{\sum_{n=0}^{\infty}}'\log\det\left[\mathbb{I} - \mathbb{N}(\kappa_{n})\right],
\end{equation}
where the primed sum runs over Matsubara frequencies $\kappa_{n} = 2\pi k_{B}T/(\hbar c)$ with the $n = 0$ term weighted by $1/2$. For a system of two objects, this matrix is $\mathbb{N} = \mathbb{T}_{1}\mathbb{U}_{12}\mathbb{T}_{2}\mathbb{U}_{21}$. Here  $\mathbb{T}_{i}$ is the T-matrix of the $i^{th}$ object, which accounts for the geometrical and material properties  of the object. $\mathbb{U}_{ij}$ are translation matrices that describe the conversion from regular waves in the coordinate system of object $i$ to outgoing waves in the system of object $j$. 

\begin{figure}[ht] 
\begin{center}
\includegraphics[width=.7\linewidth]{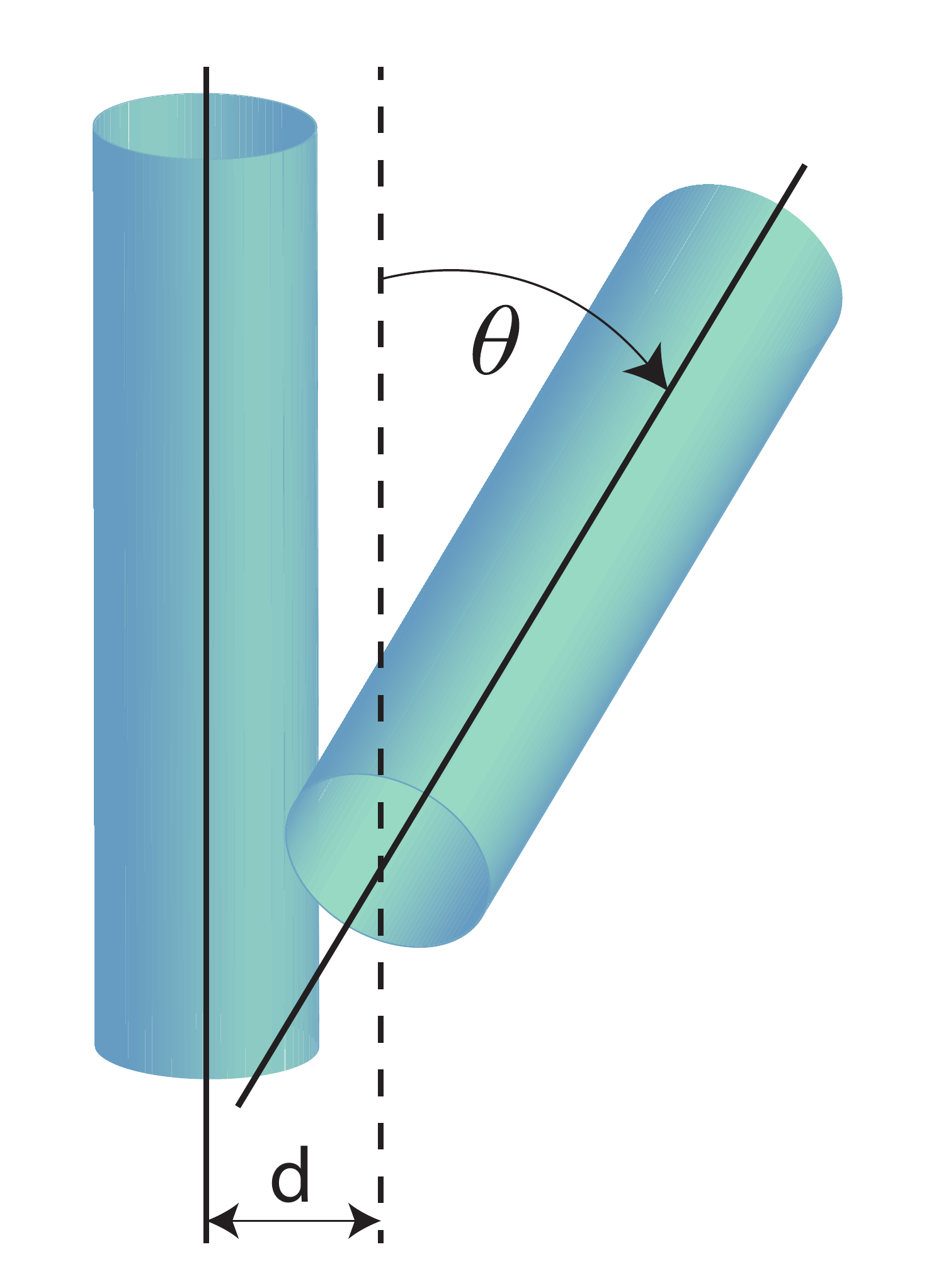} 
\caption{\label{fig:geometry}(color online) Definition of distance and orientation of two inclined cylinders with axis-to-axis separation $d$ and inclination angle $\theta$.}
\end{center}
\end{figure}

In the high temperature limit, $k_BT \gg \hbar c$, only the first Matsubara frequency contributes, and the energy can be written as
\begin{equation}\label{Energy_T_classical}
\lim_{T\to\infty}E = 
\frac{k_{B}T}{2}\log\det\left[\mathbb{I} - \mathbb{N}(0)\right]
 =-\frac{k_{B}T}{2}\sum_{p=1}^{\infty}\frac{1}{p}\text{Tr}\left[\mathbb{N}^{p}(0)\right],
\end{equation}
where the relation $\log\det A = \text{Tr}\log A $ and the expansion of $\log(1-x)$ for small $x$ have been used. The latter expression corresponds to a multiple scattering expansion since each factor of $\mathbb{N}$ describes two scattering events, one at each of the two objects. The Casimir energy at zero temperature can  be written also as a multiple scattering expansion,
\begin{equation}\label{Energy_T_0}
\begin{split}
E_0 &= \frac{\hbar c}{2\pi}\int_{0}^{\infty}d\kappa\log\det[\mathbb{I} - \mathbb{N}(\kappa)]\\
  & = - \frac{\hbar c}{2\pi}\sum_{p=1}^{\infty}\frac{1}{p}\int_{0}^{\infty}d\kappa\text{Tr}\left[\mathbb{N}^{p}(\kappa)\right] \, .
\end{split}
\end{equation}
For cylinders, the $\mathbb{T}$ matrices are generally known \cite{Review-Jamal-Emig}. The  $\mathbb{U}$ matrices
are only known for cylindrical coordinate systems with parallel axes \cite{Review-Jamal-Emig}.  For non-parallel cylinders, Eqs.~\eqref{Energy_T_classical} and \eqref{Energy_T_0} remains valid. But now the $\mathbb{U}$ matrices depend not only on the distance $d$ between the cylinder axes but also on their relative orientation, described by the inclination angle $\theta$.
In particular, we will study the case of a translation along and rotation about the $y$-axis with $\textbf{x}' = \textbf{R}_{\theta}\textbf{x} + d\hat{\textbf{y}}$ where the matrix $\textbf{R}_{\theta}$ describes a rotation about the $y$-axis by an angle $\theta$, see Fig.~\ref{fig:geometry}.

\subsection{Translation matrices}
\label{sec: 3}

In this section we derive the translation matrices $\mathbb{U}$ for scalar and electromagnetic partial waves in non-parallel cylinder coordinate systems. In particular we transform outgoing cylindrical waves with imaginary frequency $\omega=i\kappa$ to regular cylindrical wave in a coordinate system translated by a distance $d$ and rotated by an angle $\theta$.

\subsubsection{Scalar waves}

We consider regular and outgoing cylindrical wave functions which on the imaginary frequency axis are given by 
\begin{equation}\label{Energy_T_0_scalar_reg}
\phi_{n,k_{z}}^\text{reg}(\textbf{x}) = I_{n}(\rho p)e^{in\theta}e^{ik_{z}z},
\end{equation}
\begin{equation}\label{Energy_T_0_scalar_out}
\phi_{n,k_{z}}^\text{out}(\textbf{x}) = K_{n}(\rho p)e^{in\theta}e^{ik_{z}z},
\end{equation}
with $p = \sqrt{\kappa^2 + k_{z}^{2}}$ and $\rho = \sqrt{x^{2} + y^{2}}$. The outgoing waves at position $\textbf{x}' = \textbf{R}_{\theta}\textbf{x} + d\hat{\textbf{y}}$ can be expanded in terms of regular waves at $\textbf{x}$ by the linear relation
\begin{equation}\label{U_matrix_definition_crossed_cylinders}
\phi^\text{out}_{n',k'_{z}}(\textbf{x}') = \sum_{n=-\infty}^\infty\frac{L}{2\pi}\int_{-\infty}^{\infty}dk_{z}\mathbb{U}_{n'k'_{z},nk_{z}}(d,\theta)\phi_{n,k_{z}}^\text{reg}(\textbf{x}),
\end{equation}
which defines the translation matrix $\mathbb{U}_{n'k'_{z},nk_{z}}(d,\theta)$. The mathematical details of the derivation of the translation matrix are given in  App.~\ref{App. A}. For $\theta>0$, the matrix elements are given by
\begin{equation}
\label{U matriz cilindros girados theta escalar bueno}
\begin{split}
\mathbb{U}_{n'k'_{z},nk_{z}}(d,\theta) & = \frac{2\pi}{L}\frac{(-i)^{n+n'}}{\sin\theta}\left(\xi' - \sqrt{{\xi'}^{2} + 1}\right)^{n'} \\
&\times \left(\xi - \sqrt{\xi^{2} + 1}\right)^{-n}\frac{e^{- d\sqrt{{k'}_{x}^{2} + {p'}^{2}} }}{2\sqrt{{k'}_{x}^{2} + {p'}^{2}}},
\end{split}
\end{equation}
where $p'=\sqrt{\kappa^2+{k_z'}^2}$, $\xi = k_{x}/p$, $\xi' = k'_{x}/p'$, $k_{x} = (\cos(\theta)k_{z} - k'_{z})/\sin(\theta)$ and $k'_{x} = (k_{z} - \cos(\theta)k'_{z})/\sin(\theta)$.
For $\theta < 0$ the overall sign of $\mathbb{U}_{n'k'_{z},nk_{z}}(d,\theta)$ must be changed. The translation matrix elements for the inverse transformation (rotation and translation) are obtained by $d \to - d$, $\theta\to -\theta$, see App.~\ref{App. A}. When we indicate that the translation matrix corresponds to the inverse coordinate transformation by the arguments $(-d$, $-\theta)$, we have
\begin{equation}
\label{U_scalar_matrix_to_minus_hat_x}
\mathbb{U}_{n'k'_{z},nk_{z}}(-d,-\theta) = (-1)^{n+n'}\mathbb{U}_{n'k'_{z},nk_{z}}(d,\theta).
\end{equation}

\subsubsection{Electromagnetic waves}

For the EM field one expects that the translation matrix couples the polarizations
due to the different orientation of the coordinate systems. 
The outgoing vector waves are defined in terms of the outgoing scalar cylinder waves as
\begin{align}
\textbf{M}_{n,k_{z}}^\text{out} & = \frac{1}{p}\nabla\times\left(\phi_{n,k_{z}}^\text{out}\hat{\textbf{z}}\right)\, , \\
\textbf{N}_{n,k_{z}}^\text{out} & = \frac{1}{\kappa p}\nabla\times\left(\phi_{n,k_{z}}^\text{out}\hat{\textbf{z}}\right)
\end{align}
and equivalently for regular vector waves. To derive the translation matrices, we first multiply both sides of Eq. \eqref{U_matrix_definition_crossed_cylinders} by $\hat{\textbf{z}}'$ and apply $\frac{1}{p'}\nabla'\times$ on both sides of this equation. Then the left hand side gives the components of $\textbf{M}_{n,k_{z}}^\text{out}$ in the coordinate frame $\left(x',y',z'\right)$. To obtain the components in the frame $\left(x,y,z\right)$ we multiply both sides with the inverse rotation matrix ${\mathbf R}_\theta^{-1}$. This yields, using $\nabla' = {\mathbf R}_\theta\nabla$,
\begin{equation}
\label{Before_substitution}
\begin{split}
 \textbf{M}_{n',k'_{z}}^{\text{out}}(\textbf{x}')  & = {\mathbf R}_\theta^{-1}{\textbf{M}'}_{n',k'_{z}}^{\text{out}}(\textbf{x}') =   \frac{1}{p'}\sum_{n=-\infty}^\infty \frac{L}{2\pi}\int_{-\infty}^{\infty}dk_{z}\\
& \!\!\!\!\!\!\!\!\!\!\cdot \mathbb{U}_{n'k'_{z},nk_{z}}(d,\theta)\nabla\times\left[\phi^\text{reg}_{n,k_{z}}(\textbf{x})\left(\cos(\theta)\hat{\textbf{z}} - \sin(\theta)\hat{\textbf{x}}\right)\right],
\end{split}
\end{equation}
where we have used that $\hat{\textbf{z}}' = - \sin(\theta)\hat{\textbf{x}} + \cos(\theta)\hat{\textbf{z}}$. This shows that we need to express $\nabla\times\left(\phi^\text{reg}_{n,k_{z}}(\textbf{x})\hat{\textbf{x}}\right)$ in terms of $\textbf{M}_{n,k_{z}}^\text{reg}(\textbf{x})$ and $\textbf{N}_{n,k_{z}}^\text{reg}(\textbf{x})$. One can show (by expressing $\nabla$ and $\hat{\textbf{x}}$ in cylinder coordinates) that
\begin{equation}
\begin{split}
\nabla\times\left(\phi_{n,k_{z}}^\text{reg}(\textbf{x})\hat{\textbf{x}}\right) & = 
-\frac{ik_{z}}{2}\left(\textbf{M}_{n-1,k_{z}}^\text{reg}(\textbf{x}) + \textbf{M}_{n+1,k_{z}}^\text{reg}(\textbf{x})\right) \\
& + \frac{i\kappa}{2}\left(\textbf{N}_{n-1,k_{z}}^\text{reg}(\textbf{x}) - \textbf{N}_{n+1,k_{z}}^\text{reg}(\textbf{x})\right) \, .
\end{split}
\end{equation}
When this result is substituted into Eq.~\eqref{Before_substitution}, we need to shift the index $n$ in order to express the right hand side in terms of vector waves with the same index $n$. This can be done by using
\begin{equation}
\label{Subida_bajada_indice_matriz_U}
\mathbb{U}_{n'k'_{z},n\pm 1\,k_{z}}(d,\theta) = \mp i\left(\xi - \sqrt{1 + \xi^{2}}\right)^{\mp 1}\mathbb{U}_{n'k'_{z},nk_{z}}(d,\theta).
\end{equation}
This yields after some elementary algebra the translation formula for vector waves,
\begin{equation}
\label{U_em_matrix_crossed_cylinders}
\begin{split}
& \left(\begin{array}{c}
 \textbf{M}^\text{out}_{n',k'_{z}}\\
\textbf{N}^\text{out}_{n',k'_{z}}
\end{array}\right)(\textbf{x}')  = \sum_{n=-\infty}^\infty\frac{L}{2\pi}\int_{-\infty}^{\infty}dk_{z}\mathbb{U}_{n'k'_{z},nk_{z}}(d,\theta)\frac{p}{p'} \\
& \left(\begin{array}{cc}
\cos(\theta) - \sin(\theta)\frac{k_{z}}{p}\xi & \sin(\theta)\frac{\kappa}{p}\sqrt{1 + \xi^{2}}\\
-\sin(\theta)\frac{\kappa}{p}\sqrt{1 + \xi^{2}} & \cos(\theta) - \sin(\theta)\frac{k_{z}}{p}\xi
\end{array}\right)\!\!\left(\begin{array}{c}
\textbf{M}^\text{reg}_{n,k_{z}}\\
\textbf{N}^\text{reg}_{n,k_{z}}
\end{array}\right)\!(\textbf{x}) \, .
\end{split}
\end{equation}
Here we have used that $\textbf{N} = \frac{1}{\kappa}\nabla\times\textbf{M}$ and $\frac{1}{\kappa}\nabla\times\textbf{N} = - \textbf{M}$ to obtain the translation formula for $\textbf{N}$. The translation formula for the inverse coordinate transformation is given by Eq.~\eqref{U_em_matrix_crossed_cylinders} with $\mathbb{U}_{n'k'_{z},nk_{z}}$ replaced by $(-1)^{n+n'}\mathbb{U}_{n'k'_{z},nk_{z}}$, see Eq.~\eqref{U_scalar_matrix_to_minus_hat_x}. Notice that the inverted sign of $\sin\theta$ in the expression for $\textbf{z}'$ is compensated by sign changes that result from 
Eqs.~\eqref{Subida_bajada_indice_matriz_U} and \eqref{U_scalar_matrix_to_minus_hat_x}.

\subsection{Scattering amplitudes (T-matrices)}

\subsubsection{Scalar field}

For Dirichlet (D) and Neumann (N) boundary conditions, the scattering amplitudes of a cylinder of radius R are given by the expressions, 
\begin{align} 
\mathbb{T}_{n'k'_{z},nk_{z}}^{D} &= - \frac{2\pi}{L}\frac{I_{n}(pR)}{K_{n}(pR)}\delta_{n,n'}\delta(k_{z} - k'_{z}), \\ \mathbb{T}_{n'k'_{z},nk_{z}}^{N} &= - \frac{2\pi}{L}\frac{I'_{n}(pR)}{K'_{n}(pR)}\delta_{n,n'}\delta(k_{z} - k'_{z})\, , \end{align}
where $I_n$, $K_n$ are modified Bessel functions.

\subsubsection{Electromagnetic field}

For a perfect metal cylinder, the scattering amplitude 
does not couple electric (E) and magnetic (M) polarizations.
The E (M) polarization is described by D (N) boundary conditions
so that the scattering amplitude is given by
\begin{equation}
\mathbb{T}_{n'k'_{z},nk_{z}}^{EE} = \mathbb{T}_{n'k'_{z},nk_{z}}^{D},\quad 
\mathbb{T}_{n'k'_{z},nk_{z}}^{MM} =  \mathbb{T}_{n'k'_{z},nk_{z}}^{N}\, .
\end{equation}

\section{Large distance expansion}
\label{sec: 4}

In this section we obtain a large distance expansion of the scalar and electromagnetic Casimir energies at zero and high temperature. The approximations performed in this section are valid if the cylinder radius that is small compared to the distance between the cylinders. 
To perform the expansion we apply the relation $\log\det A = \text{Tr}\log A$ and  expand $\log(\mathbb{I} - \mathbb{N})\approx -\mathbb{N}$ in Eqs.~\eqref{Energy_T_classical} and \eqref{Energy_T_0}. In addition to that, we perform an expansion of the scattering amplitudes $\mathbb{T}$ in the radius $R$.

\subsection{Scalar field}

To obtain the asymptotic Casimir energy for Dirichlet boundary conditions at large distances $d \gg R$, we need to consider only the terms with $n = n' = 0$ and $p = 1$ of Eqs.~\eqref{Energy_T_classical} and \eqref{Energy_T_0}. Using for the small radius expansion of the scattering amplitude  $\mathbb{T}$ that for small $z$ one has $I_{0}(z)/K_{0}(z) \approx - \log^{-1}(z)$, we get for the energy at zero temperature
\begin{equation}\label{Scalar_Dirichlet_SRAEnergy_zero_T}
E_{0} = - \frac{\hbar c}{8 d \sin(\theta)\log^{2}\left(d/R \right)} \, .
\end{equation}
The corresponding energy for parallel cylinders of length $L\to\infty$ is \cite{Review-Jamal-Emig}
\begin{equation}
E_{0}^{\parallel} = - \frac{\hbar c L}{8\pi d^{2}\log^{2}\left(d/R\right)},
\end{equation}
so that an effective length $L_\text{eff}$ over which the inclined cylinders
interact in the limit $\theta\to 0$ can be determined by the relation
$\sin\theta = \pi d/L_\text{eff}$.

In order to study the high temperature limit $k_B T \gg \hbar c$ we need to consider only the zero Matsubara frequency contribution to the energy. A simple scaling analysis shows that $\mathbb{N}_{0\,k_{z},0\,k_{z}}(\kappa = 0)\approx 1/\abs{k_{z}}$ for $k_{z}\to 0$ with logarithmic corrections. Hence the trace of $\mathbb{N}$ is not well defined, i.e., $\mathbb{N}_{0\,k_{z},0\,k_{z}}(\kappa = 0)$ is not a trace class operator, so that $\det[\mathbb{I} - \mathbb{N}_{0\,k_{z},0\,k_{z}}(\kappa = 0)]$ is not well defined. Kenneth and Klich showed that $\mathbb{N}$ is a trace class operator for compact objects \cite{Kenneth_and_Klich}. While {\it parallel} cylinders constitute effectively a 2D problem of two compact discs, two {\it tilted} cylinders are non-compact objects whose geometry cannot be reduced to a lower dimensional one of compact objects. However, we expect the force between tilted cylinders to be well defined. This implies that the operator $\partial_{d}\mathbb{N}_{0\,k_{z},0\,k_{z}}(\kappa = 0)$ should be a trace class operator. This is indeed the case. In the high $T$ limit we obtain the force
\begin{equation}
F_{T} = - \frac{\pi k_{B}T}{4 d\sin(\theta)\log^{2}\left(d/R\right)} \, .
\end{equation}
The energy can be obtained from the force by integration, $E_{T} = \int_{d}^{\infty}d d'\,F_{T}(d')$, leading to
\begin{equation}\label{Scalar_Dirichlet_SRAEnergy_high_T}
E_{T} = -\frac{\pi k_{B}T}{4\sin(\theta)\log\left(d/R\right)}\, .
\end{equation}
This form of Casimir interaction appears to be the one with the weakest decay known to date.

For Neumann boundary conditions no logarithmic interaction appears but the energy is proportional to the product of the cross-sectional areas of the cylinders.  In Eqs.~\eqref{Energy_T_classical} and \eqref{Energy_T_0} we have to consider the terms with $p=1$, $\abs{n}\leq 1$ and $\abs{n'}\leq 1$ since $I'_{n}(z)/K'_{n}(z) = - \frac{z^{2}}{2} + \mathcal{O}\left[z^{4}\right]$ for $\abs{n}\leq 1$. With this expansion we get to lowest order in $R/d$ the energy 
at zero temperature,
\begin{equation}
 E_{0} = - \frac{\hbar c R^{4}}{320 d^{5}\sin(\theta)}[167 + \cos(2\theta)] \, , 
\end{equation} 
which can be compared to the corresponding energy for parallel cylinders \cite{Rahi-Cylinders2},
\begin{equation}
  \label{eq:D_E_parallel}
  E_0^{\|} = - \frac{7 \, \hbar c L \, R^4}{5\pi \, d^6} 
\end{equation}
so that the effective length $L_\text{eff}$ in the limit $\theta\to 0$ is determined by
$\sin \theta = (3\pi/8) d/L_\text{eff}$. In the high temperature limit the energy can be obtained directly since $\mathbb{N}(\kappa=0)$ is a trace class operator, yielding
\begin{equation}
 E_{T} = - \frac{3 \pi \, k_{B}T R^{4}}{1024 d^{4}\sin(\theta)}[98 + \cos(2\theta)] \, .  
\end{equation}
In contrast to the Dirichlet case, the energy has a more complicated orientation dependence.

\subsection{Electromagnetic field}

\begin{figure}[h] 
\begin{center}
\includegraphics[width=1.\linewidth]{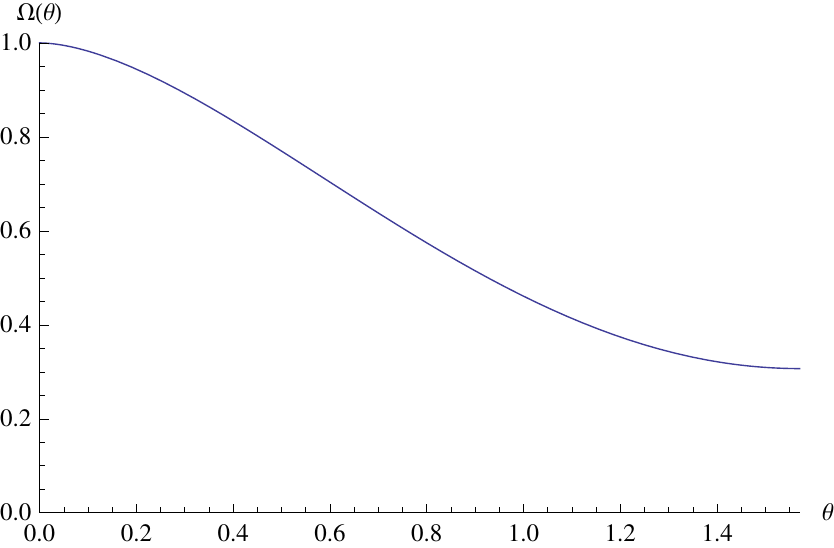} 
\caption{\label{fig:Omega} Amplitude function $\Omega(\theta)$ of the asymptotic electromagnetic Casimir energy of Eq.~(\ref{eq:Omega}) as function of inclination angle $\theta$.}
\end{center}
\end{figure}

We calculate the energies with the help of the expansions given in Eqs.~\eqref{Energy_T_classical}, \eqref{Energy_T_0}. 
In general, the translation matrix of Eq.~\eqref{U_em_matrix_crossed_cylinders} mixes the two polarizations. However,
the leading part of the energy at far distance is given by the $E$-polarized waves only due the logarithmic dependence on frequency of the scattering amplitude for $E$ modes with $n=0$ and a power-law dependence for $M$ modes. 
In the high temperature limit where all matrix elements are computed in the limit 
$\kappa \to 0$, the polarizations are decoupled at all distances since the matrix elements that couple different polarizations are proportional to $\kappa$, see Eq.~\eqref{U_em_matrix_crossed_cylinders}. We find that the asymptotic Casimir energy for perfect metal cylinders at zero temperature can be written as
\begin{equation}
\label{SRA_em_Energy_nonparallel_cylinders}
E_{0} = - \frac{\hbar c \, \Omega(\theta) }{8\, d\sin(\theta)\, \log^{2}\left(d/R\right)},
\end{equation}
where the function $\Omega(\theta)$ is defined as
\begin{widetext}
\begin{equation}
\label{eq:Omega}
\Omega(\theta) =\frac{1}{4\pi} \int_0^\pi d \vartheta \sin\vartheta \int_0^{2\pi}
d\varphi \frac{(\cos\theta\sin\vartheta +\cos\vartheta\sin\varphi \sin
  \theta)^2}
{\cos^2\varphi \sin^2\vartheta +(\cos\vartheta\sin\theta+
\sin\vartheta\sin\varphi\cos\theta)^2}   
\end{equation}
\end{widetext}
The orientation dependence described by $\Omega(\theta)$ results from the geometric factor $\cos \theta - (k_z\xi/p)\sin \theta$ that appears in the coupling of the $E$ polarization in Eq.~\eqref{U_em_matrix_crossed_cylinders}.  We have $\Omega(0) = 1$ so that for $\theta\to\,0$ the Dirichlet result given in Eq.~\eqref{Scalar_Dirichlet_SRAEnergy_zero_T} is recovered. For perpendicular cylinders one has $\Omega(\pi/2) = 1 - \log(2)$. At intermediate angles the function $\Omega(\theta)$ can be computed by numerical integration.  The corresponding result is shown in Fig.~\ref{fig:Omega}. The function $\Omega(\theta)$ can be also expanded as a Fourier series $\Omega(\theta)=\sum_{n=0}^{\infty}\Omega_{2n}\cos(2n\theta)$ with the low order coefficients given by $\Omega_{0} = 0.6137$, $\Omega_{2} = 0.3333$, $\Omega_{4} = 0.0333$ and $\Omega_{6} = 0.0096$.

Next, we consider the high temperature limit. In this limit, at all separations the polarizations are decoupled since the coupling is proportional to $\kappa$. This implies that the interaction is the sum of the energy for Dirichlet and Neumann boundary conditions. However, the presence of a ($n$-independent) geometric factor in the translation matrix elements for both $E$ and $M$ polarizations, see Eq.~\eqref{U_em_matrix_crossed_cylinders}, could modify the interaction in the electromagnetic case. That this is not the case can be seen by considering the limit $\kappa \to 0$ of the geometric factor $(p/p')[\cos \theta - (k_z\xi/p)\sin \theta]$ which is unity. Hence, in the high temperature limit, the interaction is precisely the sum of the Dirichlet and Neumann energy. At large separations, the Dirichlet contribution dominates. This leads for the electromagnetic case to the same problem as for a Dirichlet scalar field 
due to the fact that $\mathbb{N}_{0\,k_{z},0\,k_{z}}$ at $\kappa=0$ is not a trace class operator. 

However, as in the scalar case, we are able to obtain the high temperature limit of the force, because $\partial_{d}\mathbb{N}_{0\,k_{z},0\,k_{z}}(\kappa = 0)$ is a trace class operator. The far distance approximation of the force is
\begin{equation}
F_{T} = - \frac{\pi \, k_{B}T}{4 \, d\sin(\theta)\log^{2}\left(R/d\right)},
\end{equation}
and the corresponding energy is obtained by integration as
\begin{equation}
E_{T} = - \frac{\pi \, k_{B}T}{4\sin(\theta)\log\left(d/R\right)} \, ,
\end{equation}
which is identical to the Dirichlet result of Eq. \eqref{Scalar_Dirichlet_SRAEnergy_high_T}.  It should be mentioned that this slow decay with distance is a consequence of the metallic response of the cylinders.  For non-conducting cylinders a power-law decay is expected.

\section{Numerical results for smaller separations}
\label{sec: 6}

In this Section we present numerical results for the interaction between inclined cylinders.  This allows us to go beyond the limit of asymptotically large separations. We focus on the zero temperature with the energy given by Eq.~\eqref{Energy_T_0}. The broken translational symmetry of inclined cylinders leads to a matrix $\mathbb{N}$ that is non-diagonal in the continuous index $k_z$. In contrast to compact objects or systems with translational symmetry (parallel cylinders), one has to compute the determinant of a non-diagonal operator over a continuous index.  There are basically two approaches to obtain an approximation for the determinant. One can approximate the determinant by the trace of a series of powers of $\mathbb{N}$ [see Eq.~\eqref{Energy_T_0}] which is known as multiple scattering expansion or perform a discretization in $k_z$ and compute the determinant over the resulting finite set of discrete variables. While the first approach is limited to sufficiently large separations since the multiple scattering series needs to be truncated at some order, the second approach requires enough discretization points in order to minimize the discretization error. We have decided to follow the second route. This discussion shows that the evaluation of the interaction between inclined cylinders is substantially more expensive in terms of numerical operations than that of parallel cylinders. The second discrete index $n$ of the matrix $\mathbb{N}$ is truncated at some value $n_\text{max}$ as in the case of compact objects \cite{EGJK}. The choice of $n_\text{max}$ depends on the range of distances for which one would like to compute the energy. $n_\text{max}$ increases with decreasing minimal distance. For our results shown below, we found it sufficient to consider $\abs{n}\leq 3=n_\text{max}$. We studied two different values for the inclination angle, $\theta = \pi/2$ and $\theta = \pi/4$, for separations $d \geq 2.22 R$.

\begin{figure}[h] 
\begin{center}
\includegraphics[width=1.\linewidth]{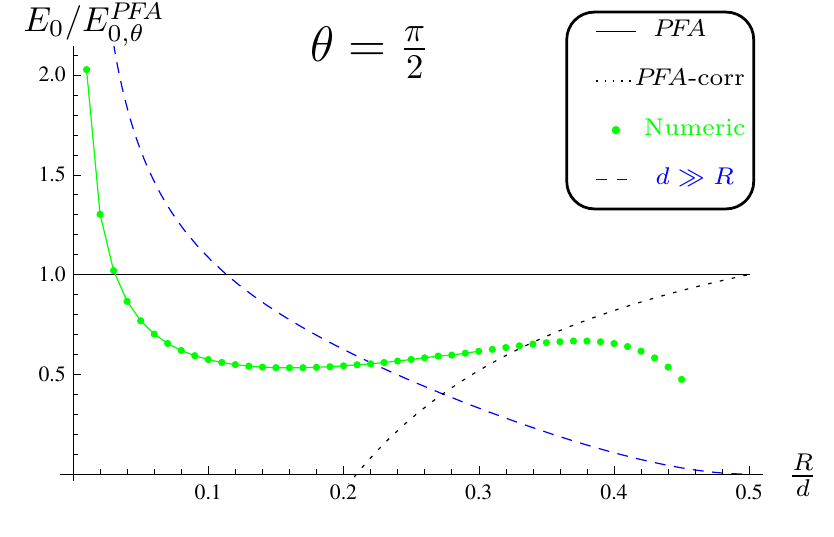} 
\caption{\label{Fig 1}(color online) Electromagnetic Casimir energy normalized to the PFA energy as function of the inverse distance for perpendicular cylinders ($\theta=\pi/2$). The dashed curve represents the asymptotic energy of Eq.~(\ref{SRA_em_Energy_nonparallel_cylinders}) and the dotted curve shows the energy of Eq.~\eqref{eq:PFA_correction}.}
\end{center}
\end{figure}

\begin{figure}[h]
\begin{center}
\includegraphics[width=1.\linewidth]{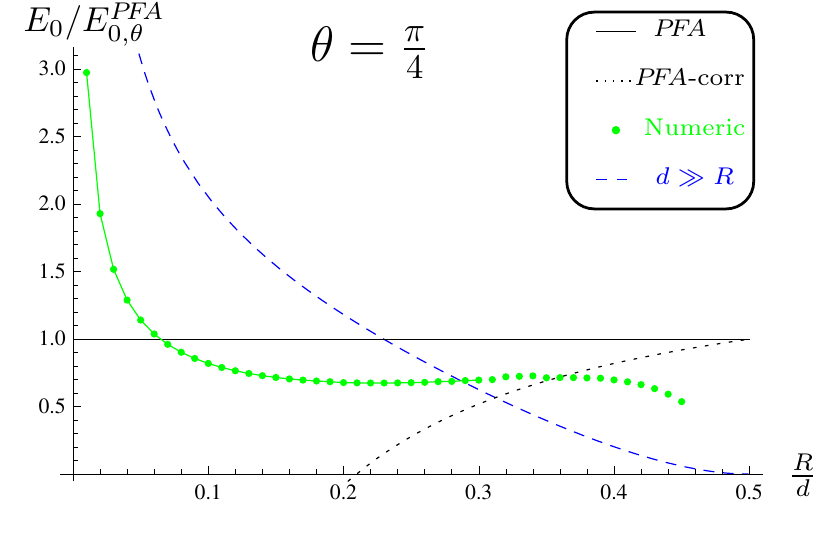} 
\caption{\label{Fig 2}(color online) Same plots as in Fig.~\ref{Fig 1} but for inclination $\theta = \pi/4$.}
\end{center}
\end{figure}

Our numerical results for the electromagnetic Casimir energies are shown in Figs.~\ref{Fig 1} and \ref{Fig 2} (dots and solid curves). They are scaled by PFA energy for cylinders with $\theta = \pi/2$ and with $\theta = \pi/4$, respectively. Regarding the approach of the numerical results to the asymptotic expression for the energy of Eq.~\eqref{SRA_em_Energy_nonparallel_cylinders} (dashed curves), we observe similar behavior as in the case of parallel cylinders \cite{Rahi-Cylinders2}: Since the energy decays logarithmically (with sub-leading logarithmic corrections), the actual energy converges to the asymptotic result only at extremely large separations that are not shown in the plots. 
At intermediate distances, higher order partial waves need to be included and are expected to lead to an approach of the energy to the PFA result at small separations. This tendency indeed can be observed for our numerical data. 
It should be noted that beyond the distance $d \approx 3.33 R$ where the curve for the numerical result starts to bend downwards, the result becomes inaccurate and more partial waves must be included. Instead of performing computations with more partial waves, we compare our numerical results to the prediction of a recently developed gradient expansion of the interaction energy for gently curved surfaces \cite{Bimonte2011-1,Bimonte2011-2}.
The latter approach yields for the first correction to PFA energy the result
\begin{equation}
  \label{eq:PFA_correction}
  E = E^\text{PFA}_{0,\theta} \left[ 1 - \frac{1}{2} \left( \frac{10}{\pi^2} - \frac{7}{24} \right) \frac{l}{R} + \ldots \right]
\end{equation}
where $l = d-2R$ is the surface-to-surface distance. This result is shown as dotted curves in Figs.~\ref{Fig 1} and \ref{Fig 2}. It can be observed that our numerical results nicely approach the prediction of Eq.~\eqref{eq:PFA_correction} for short distances. Hence, our numerical results together with Eq.~\eqref{eq:PFA_correction} provide the overall 
behavior of the Casimir interaction between inclined cylinders.


\begin{figure}[h]
\begin{center}
\includegraphics[width=1.\linewidth]{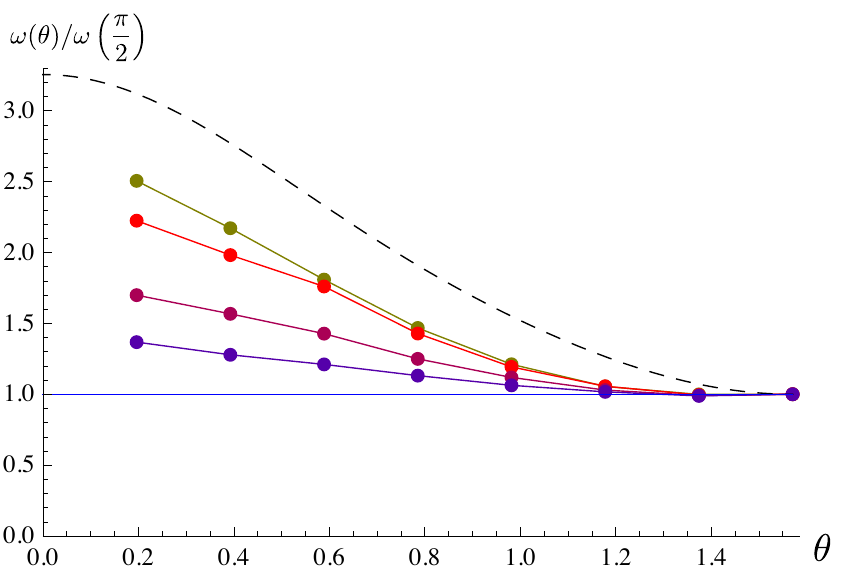} 
\caption{\label{Fig 3}(color online) Ratio $\omega(\theta)/\omega(\pi/2)$ with $\omega(\theta) = E_0(r,\theta)\sin(\theta)$ and $r=R/d$ as function of the inclination angle $\theta$. The dashed curve represents the asymptotic result of Eq.~(\ref{SRA_em_Energy_nonparallel_cylinders}). The solid curves connect numerical data for $r=0.01$, $r=0.1$, $r=0.2$, and $r=0.3$ (from top to bottom). The PFA yields  for the ratio unity.}
\end{center}
\end{figure}

We are also interested in the orientation dependence of the energy at a fixed distance. As can be seen from Eq.~\eqref{SRA_em_Energy_nonparallel_cylinders} and Eq.~\eqref{PFA_Energy_T_0_non_parallel}, the energies have a different angular dependence. To study the angular dependence of the energy at intermediate distances, we show in Fig.~\ref{Fig 3} the rescaled energy $\omega(r,\theta) = E_0(r,\theta)\sin(\theta)$ as function of the inclination angle compared to the same function at $\theta = \pi/2$, for different distances.  One observes that at large distances the dependence of $\omega(r,\theta)$ on $\theta$ is close to the asymptotic result of Eq.~(\ref{SRA_em_Energy_nonparallel_cylinders}), while for reduced distances $\omega(r,\theta)$ becomes more flat, indicating that it converges to the constant value of the PFA result.

\section{Summary and conclusion}
\label{sec: 7}

We have studied the interaction of inclined cylinders with Dirichlet, Neumann and perfect conductor boundary conditions.  In order to study this system, we have obtained the translation matrices between outgoing and regular cylindrical waves in inclined coordinate systems, both for scalar and electromagnetic waves. The interaction energies have been computed over wide range of separations, using analytic and numeric approaches. The zero temperature and classical high temperature limits were considered. In the latter, we have found that the zeroth order Matsubara term corresponds to a non-trace class operator $\mathbb{N}(\kappa=0)$ for a scalar field with Dirichlet boundary conditions and for the electromagnetic field.  However, we could obtain the energy by integration of the force which is well-defined since $\partial_{d}\mathbb{N}(\kappa=0)$ is a trace-class operator. We should remark that the zeroth order Matsubara term contributes to Casimir energy at any non zero temperature so that this result is relevant not only in the high temperature limit. The non-trace class property of $\mathbb{N}(\kappa=0)$ 
can be related to the metallic response of the cylinders. As soon as one of the two cylinders is non-conducting or one of the cylinders has a scalar field boundary condition different from the Dirichlet case, $\mathbb{N}(\kappa=0)$ is a trace class operator.

We note that our analysis for perfect metal cylinders could be easily extended to dielectric cylinders since the translation matrices remain unchanged. In this case it is expected that at large distances the energy decays according to a power-law, the zero Matsubara frequency contribution leads to a trace class operator, and the interaction is generally weaker than in the perfect metal case.

The Casimir interactions between cylinders as a function of their inclination angle $\theta$ could be experimentally probed. 
The force between cylinders with fixed inclination could be measured directly or by measuring the effect of the Casimir force on the mechanical response (oscillations) of thin metallic wires. The rather slow decay of the interaction between inclined cylinders might allow to consider separations that are well beyond the validity range of the PFA. Another experimentally interesting
quantity is the torque $\tau = - \partial E/\partial \theta$ between the cylinders for which a torsion pendulum might be employed.  For future work, it would be interesting to study the influence of finite conductivity  and the related anomalous scaling of the energy \cite{Noruzifar+2011} on the orientation dependence.

\acknowledgements 
We acknowledge helpful discussions with R.~Brito, M.~Kardar and R.~Zandi.  P.R.-L.’s research is supported by projects MOSAICO, UCM/PR34/07-15859, MODELICO (Comunidad de Madrid) and a FPU MEC grant.

\begin{appendix}

\section{Derivation of translation matrices}
\label{App. A}

In this appendix, we provide the derivation of the result in Eq.~\eqref{U matriz cilindros girados theta escalar bueno}. We start from the Fourier transform of outgoing waves  in the $xy$-plane,
\begin{eqnarray}
\label{Outgoing_2D_Fourier_Transform}
K_{n}(\rho p)e^{in\theta} & = & 2\pi(-i)^{n} \int_{-\infty}^{\infty}\frac{k_{x}}{2\pi}\int_{-\infty}^{\infty}\frac{k_{y}}{2\pi}\nonumber\\
& & \times\left(\frac{k_{x} + ik_{y}}{p}\right)^{n}\frac{e^{i(k_{x}x + k_{y}y)}}{k_{x}^{2} + k_{y}^{2} + p^{2}}.
\end{eqnarray}
The integration over $k_y$ can be easily performed using the residue theorem. Since the integrand has poles at $k_{y} = \pm i\sqrt{p^{2} + k_{x}^{2}}$, we obtain
\begin{eqnarray}
K_{n}(\rho p)e^{in\theta} & = & (-i)^{n} \int_{-\infty}^{\infty}dk_{x}\left(\frac{k_{x} \mp \sqrt{k_{x}^{2} + p^{2}}}{p}\right)^{n}\nonumber\\
& & \times\frac{e^{ik_{x}x \mp y\sqrt{k_{x}^{2} + p^{2}} }}{2\sqrt{k_{x}^{2} + p^{2}}},
\end{eqnarray}
where the minus (plus) sign applies to $y > 0$ ($y < 0$). 
With the coordinate transformation  $\textbf{x}' = \textbf{R}_{\theta}\textbf{x} + d\hat{\textbf{y}}$ and the corresponding wave
vector components $k_x=k_x' \cos \theta - k_z' \sin \theta$,
$k_y'=k_y=\pm i \sqrt{\kappa^2+k_x^2+k_z^2}$, $k_z=k_x'\sin\theta
+k_z' \cos\theta$ we get the outgoing wave in the primed coordinate system for $y'>0$ ($y'<0$)
\begin{eqnarray}
\label{eq:app_phi_out}
\phi_{n',k'_{z}}^{out}(\textbf{x}') & = & (-i)^{n'}\int_{-\infty}^{\infty}dk'_{x}\left(\frac{k_{x}' \mp \sqrt{{k_{x}'}^{2} + {p'}^{2}}}{p'}\right)^{n'}\nonumber\\
& & \times\frac{e^{i(k_{x}x +  k_{y}y +k_{z}z)} e^{\mp  \sqrt{{k_x'}^{2} + {p'}^{2}} d}}{2\sqrt{{k_x'}^{2} + {p'}^{2}}},
\end{eqnarray}
with $p'=\sqrt{\kappa^2+{k_z'}^2}$. Now we use the expansion of 2D plane waves in cylindrical waves,
\begin{eqnarray}
e^{i(xk_{z} + k_{y}y)} & = & \sum_{n\in\mathbb{Z}}i^{n}J_{n}\left(\rho\sqrt{k_{x}^{2} + k_{y}^{2}}\right)e^{in\theta}\nonumber\\
& & \times e^{-in\arccos\left(\frac{k_{x}}{\sqrt{k_{x}^{2} + k_{y}^{2}}}\right)}\, .
\end{eqnarray}
With the relations $ip = \sqrt{k_{x}^{2} + k_{y}^{2}}$ and
\begin{equation}
e^{-in\arccos\left(\frac{k_{x}}{\sqrt{k_{x}^{2} + k_{y}^{2}}}\right)} = i^{n}\left(\frac{k_{x} + ik_{y}}{p}\right)^{-n}\, , 
\end{equation}
the plane wave can be written as
\begin{equation}
e^{i(xk_{z} + k_{y}y)} = \sum_{n\in\mathbb{Z}}(-i)^{n}I_{n}\left(\rho p\right)e^{in\theta}\left(\frac{k_{x} - \sqrt{k_{x}^{2} + p^{2}}}{p}\right)^{-n}.
\end{equation}
By insertion of this sum in Eq.~\eqref{eq:app_phi_out}, we can express the outgoing wave in terms of incoming waves,
\begin{widetext}
\begin{equation}
\label{U_matrix_before_change_from_ky_to_kz}
\phi_{n',k'_{z}}^{out}(\textbf{x}') = \sum_{n\in\mathbb{Z}}(-i)^{n+n'}\int_{-\infty}^{\infty}dk'_{x}\left(\frac{k'_{x} - \sqrt{{k'}_{x}^{2} + {p'}^{2}}}{p'}\right)^{n'}\left(\frac{k_{x} - \sqrt{k_{x}^{2} + p^{2}}}{p}\right)^{-n}\frac{e^{- d\sqrt{{k'}_{x}^{2} + {p'}^{2}} }}{2\sqrt{{k'}_{x}^{2} + {p'}^{2}}}I_{n}\left(\rho p\right)e^{in\theta}e^{ik_{z}z} \, ,
\end{equation}
where we assumed $y'>0$, $d>0$, and $\theta>0$.
In order to bring the result into the form of Eq.~\eqref{U_matrix_definition_crossed_cylinders}, we change the variable of integration from $k_{x}'$ to $k_{z}$ using $k_{z} = \cos(\theta)k'_{z} + \sin(\theta)k'_{x}$ and  $dk'_{x} = dk_{z}/\sin\theta$, which yields
\begin{equation}
\label{eq:app_phi_out_2}
\phi_{n',k'_{z}}^{out}(\textbf{x}') = \sum_{n\in\mathbb{Z}} \int_{-\infty}^{\infty}dk_{z}\frac{(-i)^{n + n'}}{\sin(\theta)}\left(\frac{k'_{x} - \sqrt{{k'}_{x}^{2} + {p'}^{2}}}{p'}\right)^{n'}\left(\frac{k_{x} - \sqrt{k_{x}^{2} + p^{2}}}{p}\right)^{-n}\frac{e^{- d\sqrt{{k'}_{x}^{2} + {p'}^{2}} }}{2\sqrt{{k'}_{x}^{2} + {p'}^{2}}}\phi_{n,k_{z}}^{reg}(\textbf{x}),
\end{equation}
\end{widetext}
where we have used $\phi_{n,k_{z}}^{reg}(\textbf{x}) = I_{n}\left(\rho p\right)e^{in\theta}e^{ik_{z}z}$, and
it is understood that 
\begin{equation}
    k_x = k_z \cot\theta -\frac{k_z'}{\sin\theta} , \quad
   k_x' = \frac{k_z}{\sin\theta} - k_z'  \cot\theta \, .
\end{equation}
With the definitions $\xi = k_{x}/p$ and $\xi' = k'_{x}/p'$ Eq.~\eqref{eq:app_phi_out_2} is equivalent to Eq.~\eqref{U matriz cilindros girados theta escalar bueno}.

For $\theta < 0$ the overall sign of $\mathbb{U}_{n'k'_{z},nk_{z}}(d,\theta)$ must be changed. The translation matrix elements for the inverse of the transformation $\textbf{x}' = \textbf{R}_{\theta}\textbf{x} + d\hat{\textbf{y}}$ is obtained by assuming $y' < 0$ and $d \to - d$, $\theta\to -\theta$. When we indicate by the arguments $(-d, -\theta)$ that the translation matrix corresponds to the inverse coordinate transformation, we have
\begin{equation}
\mathbb{U}_{n'k'_{z},nk_{z}}(-d,-\theta) = (-1)^{n+n'}\mathbb{U}_{n'k'_{z},nk_{z}}(d,\theta) \, .
\end{equation}
Finally, we note that the translation matrices for parallel cylindrical coordinate systems follow from a direct integration of Eq.~\eqref{U_matrix_before_change_from_ky_to_kz} over $k'_{x}$ in the limit $\theta\to 0$.

\section{Proximity Force Approximation}
\label{App. B}

In this appendix we derive the PFA for the energy of two inclined cylinders at zero temperature, Eq.~\eqref{PFA_Energy_T_0_non_parallel}, and in the high temperature limit, Eq.~\eqref{PFA_Energy_T_cl_non_parallel}.
The area across which the two inclined cylinders overlap, viewed along the axis that is perpendicular to the two cylinder axes and that intersects the axes in their crossing point, forms a parallelogram of edge length $2R/\sin\theta$. Let us denote the coordinates along the edges of this parallelogram as $u$ and $v$. Then the local distance $h(u,v)$ between the two cylinder surfaces, measured normal to the plane that is spanned by the cylinder axes, is given by the function
\begin{eqnarray}
h(u,v) & = & d - \sqrt{R^{2} - \left(u\sin\theta - R\right)^{2}}\nonumber\\
& & - \sqrt{R^{2} - \left(v\sin\theta - R\right)^{2}}\, .
\end{eqnarray}
where $d$ is the distance between the cylinder axes. Taking into account that a surface element of the parallelogram is given by $\sin(\theta) \,dudv$, the PFA energy at zero temperature can be written as
\begin{equation}
E^{PFA}_{0,\theta} = - \frac{\hbar c\, \pi^{2}}{720}\int_{0}^{\frac{2R}{\sin(\theta)}}\int_{0}^{\frac{2R}{\sin(\theta)}}\frac{\sin(\theta)dudv}{h^{3}(u,v)} \, .
\end{equation}
When we introduce the surface-to-surface distance $l=d-2R$, after performing a change of integration variables the
energy can be written as
\begin{equation}
  \label{Exact_PFA}
\begin{split}
 &  E^{PFA}_{0,\theta} = -\frac{\pi^2\, \hbar c}{720} \frac{1}{\sin\theta}
\frac{l}{R^2} \\
& \times \int_{-\sqrt{R/l}}^{\sqrt{R/l}} \int_{-\sqrt{R/l}}^{\sqrt{R/l}}  \frac{ds
  dt}{\left[\frac{l}{R}+2-\sqrt{1-\frac{l}{R}s^2}-\sqrt{1-\frac{l}{R}t^2}\right]^3}  \, .
\end{split}
\end{equation}
This expression has the advantage that we can expand the square roots
for small $l/R$ which leads to 
\begin{eqnarray}
E^{PFA}_{0,\theta} & = & - \frac{\pi^{2} \, \hbar c }{720}\frac{1}{\sin(\theta)}\frac{R}{l^{2}}\nonumber\\
& & \times\int_{-\sqrt{R/l}}^{\sqrt{R/l}}\int_{-\sqrt{R/l}}^{\sqrt{R/l}}\frac{dsdt}{\left[1 + \frac{1}{2}\left(s^{2} + t^{2}\right)\right]^{3}} \, .
\end{eqnarray}
In this expression we can extend the integration limits to infinity to obtain the limiting behavior for small $l/R$. This yields
\begin{eqnarray}\label{PFA_em_Energy_nonparallel_cylinders_zero_T}
\lim_{\frac{l}{R}\to\,0}E^{PFA}_{0,\theta} & = & - \frac{\pi^{2}\, \hbar c}{720}\frac{1}{\sin(\theta)}\frac{R}{l^{2}}\int_{0}^{2\pi}d\varphi\int_{0}^{\infty}\frac{\rho d\rho}{\left(1 + \frac{\rho^{2}}{2}\right)^{3}}\nonumber\\
& = & - \frac{\pi^{3}\, \hbar c}{720}\frac{1}{\sin(\theta)}\frac{R}{l^{2}} \, ,
\end{eqnarray}
which is the result of Eq.~\eqref{PFA_Energy_T_0_non_parallel}.
For small $l/R$ the latter  approximation deviates from the exact integral of Eq.~\eqref{Exact_PFA} by less than $1\%$ for $l/R < 0.01$. It is instructive to compare this PFA energy to the one for two spheres of radius $R$ and surface-to-surface distance $l$ which is 
\begin{equation}
  \label{eq:PFA_2_spheres}
E^{PFA}_\text{spheres} = - \frac{\pi^{3}\, \hbar c}{1440}\frac{R}{l^{2}}  
\end{equation}
and hence reduced by a factor of $1/2$ compared to the case of perpendicular cylinders with $\theta = \pi/2$.

In analogy to the above computation, it is possible to obtain the PFA energy for the high temperature limit.
Using the high temperature interaction between two perfectly reflecting plates, the PFA energy for two inclined cylinders at high temperatures is given by 
\begin{equation} E^{PFA}_{cl,\theta} = - k_{B}T\, \frac{\zeta(3)}{8\pi}\int_{0}^{\frac{2R}{\sin(\theta)}}\int_{0}^{\frac{2R}{\sin(\theta)}}\frac{\sin(\theta)dudv}{h^{2}(u,v)}\, .  \end{equation} 
Performing the same transformations and approximation as for $T=0$, we find for small $l/R$ the PFA energy 
\begin{eqnarray}
  \lim_{\frac{l}{R}\to\,0}E_{PFA}^{cl} & = & - k_{B}T\, \frac{\zeta(3)}{8\pi}\frac{1}{\sin(\theta)}\frac{R}{l}\int_{0}^{2\pi}d\varphi\int_{0}^{\infty}\hspace{-10pt}\frac{\rho d\rho}{\left(1 + \frac{\rho^{2}}{2}\right)^{2}}\nonumber\\
  & = & - k_{B}T\, \frac{\zeta(3)}{4}\frac{1}{\sin(\theta)}\frac{R}{l} \, ,
\end{eqnarray}
which corresponds to Eq.~\eqref{PFA_Energy_T_cl_non_parallel}.

\end{appendix}

\bibliographystyle{apsrev}
\bibliography{tilted_cylinders}

\end{document}